\documentclass[12pt,aps,pra,showpacs,tightenlines]{revtex4}
\usepackage{graphicx}

\begin{document}

\title{Huygens' construction in a dispersive medium \\moving at a constant
velocity}

\author{Aleksandar Gjurchinovski}

\email{agjurcin@iunona.pmf.ukim.edu.mk}

\affiliation{Department of Physics, Faculty of Natural Sciences 
and Mathematics, Sts.\ Cyril and Methodius University,
P.\ O.\ Box 162, 1000 Skopje, Macedonia}

\author{Kostadin Tren\v cevski}

\email{kostatre@iunona.pmf.ukim.edu.mk}

\affiliation{Department of Mathematics, Faculty of Natural Sciences 
and Mathematics, Sts.\ Cyril and Methodius University,
P.\ O.\ Box 162, 1000 Skopje, Macedonia}

\begin{abstract}
We extend the method of Huygens' construction in a uniformly moving optical medium
[Am. J. Phys. {\bf 72}, 934 (2004)] to the case when the medium is dispersive 
in its rest frame of reference. The first-order Huygens' construction analysis of the
light drag in a transversely moving dispersive slab is in agreement with the 
results of the experiment by Jones (1975) and with the Player-Rogers formula 
for the downstream deflection of the beam. 
The derivation purports that the original Huygens' principle remains valid in 
non-stationary situations if it is modified to include the relativistic 
effects on the secondary wavelets caused by the motion of the medium. 
\end{abstract}

\pacs{42.15.-i, 03.30.+p}

\date{October 20, 2007}

\maketitle

\section{Introduction}

In recent paper \cite{gjurchinovski}, one of us (A.G.) presented a 
method for calculating the deflection of light refracted from a 
non-dispersive homogeneous optical material moving at a constant 
velocity parallel to the interface.
The method was based on Huygens' construction applied to the secondary 
wavelets, which were shown to be distorted, elliptic-shaped dragged 
ovals as a consequence of the motion of the medium. The obtained 
refraction law was used to explain the measurements for the transverse 
displacement of the light-beam probe in the experiment by Jones (1971-2), 
in which the beam was allowed to pass through a uniformly rotating disk 
made of a non-dispersive glass parallel to the axis of rotation \cite{jones1,jones2}.
The agreement of the refraction formula with the results of the experiment 
and with the formula for the transverse Fresnel-Fizeau light drag
confirmed the validity of Huygens' construction as a ray-tracing tool
in a dispersionless optical medium in uniform rectilinear motion.

In the same paper, it was noted that the rotating disk experiment was 
repeated by the same Jones a few years later (1975) using a highly 
dispersive glass material \cite{jones3}. The deflection of the probe was significantly 
enhanced, and the value of the displacement was found to be different 
from the one predicted by the formula developed for the non-dispersive 
situation. It has been shown by Player \cite{player} and Rogers \cite{rogers} that the original
formula for the transverse Fresnel-Fizeau deflection should be modified 
by the presence of an additional term due to dispersion, in agreement
with the results of the repeated Jones experiment. 

The purpose of the present paper is to extend the method of Huygens' 
construction in a uniformly moving optical medium to the case when the
medium in question is dispersive in its rest frame of reference.
In Sec. II we investigate the shape of a Huygens' wavelet in the 
presence of a dispersive medium moving at a constant velocity, and show that 
the shape of the wavelet is much more complicated than for a non-dispersive 
situation. We use these arguments in Sec. III to trace the advancement 
of a plane-polarized light beam normally incident upon a uniformly moving 
material slab, by applying the Huygens' construction to the deformed secondary 
wavelets. By limiting the analysis when the speed of the slab is much less than the 
speed of light in vacuum, we obtain the Player-Rogers formula for the transverse 
displacement of the beam.

\section{The shape of the secondary wavelets in a uniformly moving 
dispersive medium}

For the sake of simplicity, we will treat the 
problem two-dimensionally ($z=z'=0$), although the same line of reasoning is valid
in a more general three-dimensional case. Consider an observer in 
$S'$-frame, and with respect to him -- a homogeneous, isotropic, and transparent 
optical material at rest. The $S'$-observer measures the phase speed of light 
in the medium to be $c/n(\omega')$, which is a function of the locally 
measured frequency of the wave $\omega'$. Here, $n(\omega')$ is the phase 
refractive index, and $c$ is the speed of light in vacuum. With respect to $S'$, 
the space-time evolution of a secondary wavelet emanating from a given point in 
space is described by:
\begin{equation}
x'^2+y'^2=\left[c/n(\omega')\right]^2t'^2,
\end{equation}
where we choose the origin of the $x'y'$ coordinate system to overlap with
the origin of the wavelet $(t'=0)$. By Lorentz-transforming Eq. (1), we find the
shape of the elementary wavefront with respect to an observer in $S$-frame 
to whom the medium is moving at a constant speed $u$ in the positive direction
of the $x$-axis:
\begin{equation}
(x-ut)^2+y^2(1-u^2/c^2)=n(\omega')^{-2}(ct-ux/c)^2.
\end{equation}
But, the observer in $S$-frame uses $\omega$ instead of $\omega'$, and thus
measures $n(\omega)$ instead of $n(\omega')$.
To him, the frequency of the wave $\omega$ is not equal in every direction as 
for the $S'$-observer, but will vary with the angle $\varphi$ between the 
direction of the ray and the velocity of the medium. The transition from
$n(\omega')$ to $n(\omega)$ can be accomplished by using the Doppler formula:
\begin{equation}
\omega'=\omega\ {1-\left[un(\omega)/c\right]\cos\varphi \over\sqrt{1-u^2/c^2}},
\end{equation}
from which, to the first order in $u/c$, we have:
\begin{equation}
n(\omega')=n(\omega-\omega un(\omega)\cos\varphi/c)\approx 
n(\omega)\left(1-{\omega u\over c}{\partial n(\omega)\over\partial\omega}
\cos\varphi\right).	
\end{equation}
Expressing the angle $\varphi$ via the point of measurement $(x,y)$
of the wave characteristics, we have:
\begin{equation}
n(\omega')\approx n(\omega)\left(1-{\omega u\over c}{\partial 
	n(\omega)\over\partial\omega}{x\over\sqrt{x^2+y^2}}\right).	
\end{equation}
We conclude that the shape of the elementary light pulse in $S$-frame
described by Eq. (2) is generally a complicated curve, not necessarily an ellipse 
as in the case of a uniformly moving non-dispersive material.

\section{Derivation of the Player-Rogers formula}

We will investigate the deflection of light from a dispersive material 
slab moving at a constant speed $u$ parallel to its surface (see Fig. 1). 
We consider the speed $u$ of the slab to be much less than $c$, and take 
the incident light to be normal to the velocity of the slab, therefore 
resembling the Jones' setup. As a result of the motion of the slab, 
the incident beam will undergo a continual deflection inside the 
moving medium in the direction of its motion, and will eventually 
emerge from the slab displaced at a distance $q$ parallel 
to its original direction before the entrance. To find the displacement
$q$ of the beam, we will use Huygens' construction on the distorted 
secondary wavelets.
\begin{figure}
\includegraphics[width=.70\textwidth,height=!]{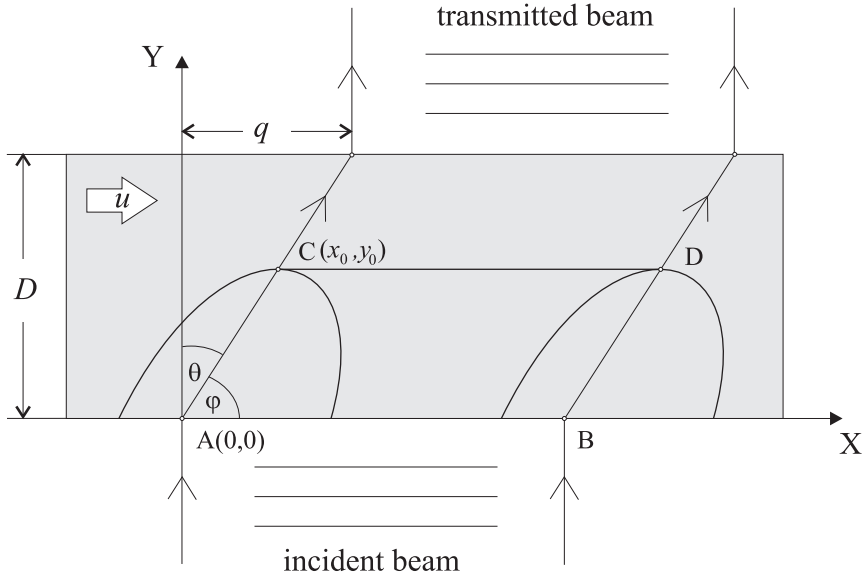}
\caption{Huygens' construction of the refracted wavefront $CD$
in the material slab that moves at a constant speed $u$ in the direction of
the positive $x$-axis. The motion of the slab is causing 
the distorsion of the elementary wavefronts originating along $AB$.
The size of the wavelets are exaggerated for convenience.}
\end{figure} 
At the instant when the incident wavefront $AB$ reaches the slab, the points along 
the interface will start radiating secondary wavelets in accord with Eq. (2). 
The envelope $CD$ of these wavelets forms the wavefront of the light 
beam at a later time $t$ counted from the beginning of the radiation of 
the elementary sources along the interface. The wavefront $CD$ is a 
tangent line of all the elementary wavefronts, and it touches the 
elementary wavefront emanating from the initially disturbed point $A$ 
at the point $C(x_0,y_0)$. The deflection angle $\theta$ of the beam is: 
\begin{equation}
\tan\theta={x_0\over y_0},
\end{equation}
Since the slope of the tangent line $CD$ is zero, we can find $x_0$ and $y_0$ 
by implicit differentiation of Eq. (2) with respect to $x$ and equating 
the terms $dy/dx$ to zero for the point $(x_0,y_0)$ in question. Hence, 
we obtain:
\begin{equation}
x_0-ut=-{u\over c}n(\omega')^{-2}(ct-ux_0/c)-n(\omega')^{-3}(ct-ux_0/c)^2
\left.{\partial n(\omega')\over\partial x}\right|_{x_0,y_0},
\end{equation}
where we have taken into account that $n(\omega')$ is a function on $x$ and $y$ 
according to Eq. (5). From Eq. (5), we obtain:
\begin{eqnarray}
n(\omega')^{-1}&\approx& n(\omega)^{-1}\left(1+{\omega u\over c}{\partial 
	n(\omega)\over\partial\omega}{x_0\over\sqrt{x_0^2+y_0^2}}\right), \\
\left.{\partial n(\omega')\over\partial x}\right|_{x_0,y_0}&\approx& -{\omega u\over c}n(\omega)
{\partial n(\omega)\over\partial\omega}{y_0^2\over(x_0^2+y_0^2)^{3/2}}.
\end{eqnarray}
By substitution of Eqs. (8) and (9) into Eq. (7), and neglecting the second 
and higher order terms in $u/c$, we obtain:
\begin{equation}
x_0\approx ut\left(1-{1\over n(\omega)^2}\right)+{\omega cut^2\over n(\omega)^2}
{\partial n(\omega)\over\partial\omega}{1\over y_0},
\end{equation}
where we used the approximations $y_0^2(x_0^2+y_0^2)^{-3/2}\approx y_0^{-1}$ 
and $x_0(x_0^2+y_0^2)^{-1/2}\approx x_0/y_0\rightarrow0$ from the fact that $x_0\ll y_0$ 
when $u\ll c$. Since the point of tangency $C(x_0,y_0)$ is a solution of Eq. (2), 
we have:
\begin{equation}
y_0=\left({n(\omega')^{-2}(ct-ux_0/c)^2-(x_0-ut)^2\over (1-u^2/c^2)}\right)^{1/2},
\end{equation}
which in the limit $x_0\rightarrow0$ reduces to:
\begin{equation}
y_0\approx{ct\over n(\omega)}.
\end{equation}
By putting Eqs. (10) and (12) for $x_0$ and $y_0$ 
into Eq. (6), we have: 
\begin{equation}
\tan\theta\approx {u\over c}\left(n(\omega)-{1\over n(\omega)}+\omega
{\partial n(\omega)\over\partial\omega}\right),
\end{equation}
which is an expression for the deflection angle $\theta$ of the beam to the 
first order in $u/c$. Taking into account that $\tan\theta=q/D$, we obtain 
the Player-Rogers formula for the transverse displacement $q$ of the beam:
\begin{equation}
q\approx D{u\over c}\left(n(\omega)-{1\over n(\omega)}+\omega
{\partial n(\omega)\over\partial\omega}\right),
\end{equation}
where $D$ is the thickness of the slab.

\section{Concluding remarks}

The theoretical and experimental investigations of light propagation 
in moving media are very topical, leading to a variety of novel and exotic
effects in relativistic and quantum optics 
\cite{artoni1,carusotto1,artoni2,artoni3,carusotto2,leonhardt1,leonhardt2,
leonhardt3,leonhardt4,leonhardt5}.  
In this and the preceding paper \cite{gjurchinovski} we have limited our discussion to a
uniformly moving media and show that Huygens' construction can be used
to trace the path of the beam inside the moving medium if we take into
account that the secondary wavelets are distorted as a consequence of
the motion of the medium. The obtained formulas for the deflection of the
beam to the first order in $u/c$ coincide with the Fresnel and the 
Player-Rogers formulas for the transverse drag in the non-dispersive and
dispersive case, respectively, in agreement with the experiments by 
Jones (1971-75). 

The described Huygens-construction analysis of light propagation in 
uniformly moving media can have certain implications into the standard textbook 
discussion of Huygens-Fresnel principle. It offers a simple geometrical 
method of approach in the framework of introductory physics courses that 
might bring the subject of optics of uniformly moving media into the typical 
undergraduate classroom.


\begin{thebibliography}{99}

\bibitem{gjurchinovski}
A. Gjurchinovski, Am. J. Phys. {\bf 72}, 934 (2004).

\bibitem{jones1}
R. V. Jones, J. Phys. A {\bf4}, L1 (1971).

\bibitem{jones2}
R. V. Jones, Proc. R. Soc. Lond. A {\bf 328}, 337 (1972).

\bibitem{jones3}
R. V. Jones, Proc. R. Soc. Lond. A {\bf 345}, 351 (1975).

\bibitem{player}
M. A. Player, Proc. R. Soc. Lond. A {\bf 345}, 343 (1975).

\bibitem{rogers}
G. L. Rogers, Proc. R. Soc. Lond. A {\bf 345}, 345 (1975).

\bibitem{artoni1}
M. Artoni, I. Carusotto, G. C. La Rocca and F. Bassani, 
Phys. Rev. Lett. {\bf 86}, 2549 (2001).

\bibitem{carusotto1}
I. Carusotto, M. Artoni, G. C. La Rocca and F. Bassani, 
Phys. Rev. Lett. {\bf 87}, 064801 (2001).

\bibitem{artoni2}
M. Artoni, I. Carusotto, G. C. La Rocca and F. Bassani, 
J. Opt. B: Quantum Semiclass. Opt. {\bf 4}, S345 (2002).

\bibitem{artoni3}
M. Artoni, I. Carusotto, Phys. Rev. A {\bf 67}, 011602(R) (2003).

\bibitem{carusotto2}
I. Carusotto, M. Artoni, G. C. La Rocca and F. Bassani, 
Phys. Rev. A {\bf 68}, 063819 (2003).

\bibitem{leonhardt1}
U. Leonhardt and P. Piwnicki, Phys. Rev. A {\bf 60}, 4301 (1999).

\bibitem{leonhardt2}
U. Leonhardt and P. Piwnicki, Contemp. Phys. {\bf 41}, 301 (2000).

\bibitem{leonhardt3}
U. Leonhardt and P. Piwnicki, J. Mod. Opt. {\bf 48}, 977 (2001). 

\bibitem{leonhardt4}
U. Leonhardt, Nature (London) {\bf 415}, 406 (2002).

\bibitem{leonhardt5}
U. Leonhardt and T. G. Philbin, New J. Phys. {\bf 8}, 247 (2006).


\end{thebibliography}
\end{document}